\documentclass[twocolumn,aps,prl,showpacs, amsmath,amssymb]{revtex4}

\usepackage{graphicx}
\usepackage{dcolumn}
\usepackage{bm}

\begin{document}

\title{Clustering instability in a freely falling granular jet}
\author{Matthias E. M\"{o}bius}
\affiliation{The James Franck Institute and Department of Physics,
The University of Chicago, Chicago, Illinois 60637}

\date{\today}

\begin{abstract}
This paper investigates a clustering instability of a freely falling
granular jet composed of $100\mu\textnormal{m}$ glass spheres. The
granular flow out of a circular nozzle starts out spatially uniform
and then, further downstream, breaks up into well defined clusters.
The role of air is investigated in this phenomenon by changing the
ambient air pressure down to $1/5000^{\textnormal{th}}$ atm. An
optical method is used that measures inhomogeneities in the flow in
order to quantify the growth of the clusters. Clustering is observed
down to the lowest pressure and the presence of air leads to larger
drops but does not initiate the drop formation. The analysis shows
that the drop size is set by fluctuations on the order of the size
of the particles at the nozzle.
\end{abstract}

\maketitle

\section{I. Introduction}

\begin{figure*}
\begin{center}
\includegraphics[width=7in]{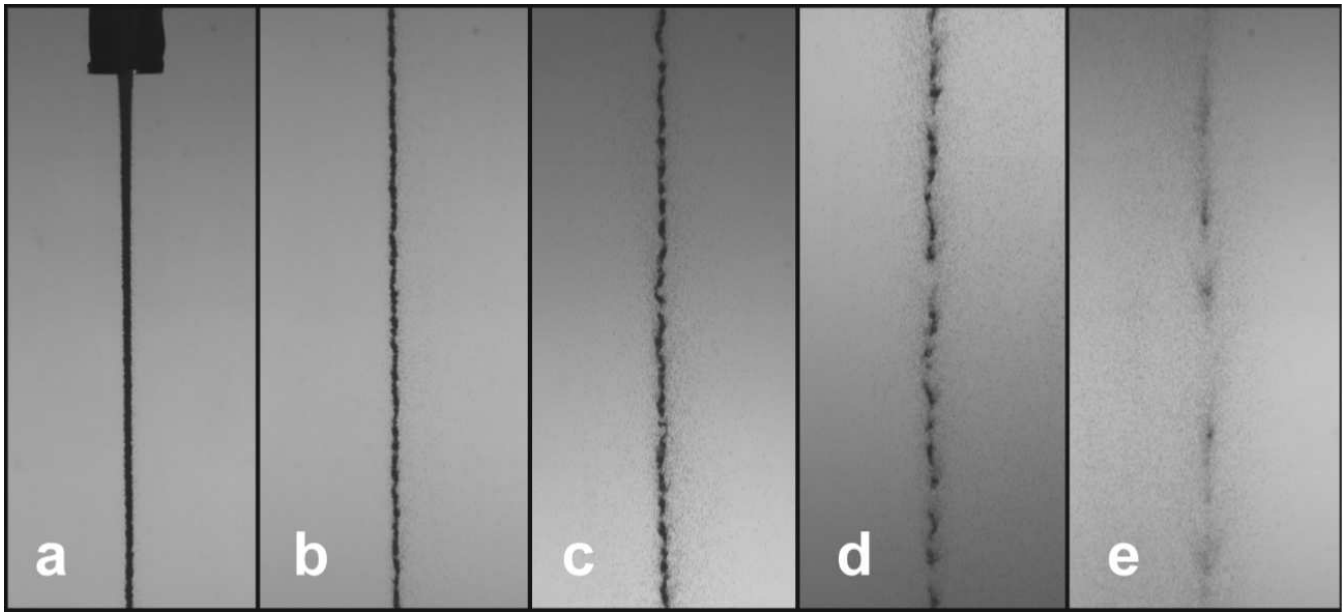}
\caption{\label{100micron} $100$ micron glass spheres draining from
the $4$ mm wide circular nozzle at atmospheric pressure. This porous
funnel is the same as used in the experiment. Pictures are taken at
different depths $z$ below the nozzle. (a) at the nozzle; (b) $z=25$
cm; (c) $z=55$ cm; (d) $z=75$ cm; (e) $z=130$ cm.}
\end{center}
\end{figure*}

Granular media often appear to behave like ordinary fluids. One can
pour sand into a bucket or use it in hour glasses. This paper
investigates how a freely falling granular jet emanating from an
aperture becomes inhomogeneous and starts to form droplets (Fig.
\ref{100micron}) similar to an ordinary fluid column breaking up due
to the Rayleigh-Plateau instability \cite{eggers}. Despite this
apparent similarity in behavior there are considerable differences
between fluid and granular flows. The fluid jet instability is
driven by the surface tension of the liquid. Dry, non-cohesive
granular media, however, do not possess surface tension, so it is
surprising to observe clustering in a granular jet.

Due to the lack of surface tension, the clustering instability must
be driven by something else. In general, inhomogeneities in granular
flows are quite common due to friction between particles and
geometrical constraints imposed by boundaries. This can lead to
arching and jamming. In the case studied here, however, the clusters
form {\it in the absence} of boundaries. This distinguishes it from
other clustering phenomena in granular flows such as density waves
in funnels and vertical pipes
\cite{baxter,schick,bertho,raafat,veje1,veje2}.

The clustering of freely falling granular jets appears to be
generic. Recent experiments revealed that after the impact of a
large sphere on a loosely packed bed of small particles a
surprisingly tall granular jet emerges \cite{thoroddsen, lohse,
royer}. In this case as well the jet breaks up into droplets whose
diameter is comparable to the jet diameter. This is another example
of drop formation in the absence of nearby boundaries.

This study investigates, as a function of air pressure, drop
formation from the flow out of a funnel. Air is important when
dealing with small grains. Viscous drag can easily exceed the
particle's weight for the $100$ $\mu$m glass spheres used in this
experiment. Moreover, investigating the role of air enables a
comparison with recent experiments on drop formation in an
underwater granular jet \cite{schaflinger,nicolas}. In my experiment
I can tune the influence of the surrounding "liquid" - namely the
air. Studying the effect of air in this phenomenon is important for
pinpointing the clustering mechanism. In particular, I study the
properties of the drops, such as their size and growth as they
freely fall, as a function of pressure.

The typical drop size decreases with decreasing pressure down to
some threshold pressure below which the size remains unchanged.
Thus, air influences the clustering but is not necessary for
initiating the process. At all pressures the drops grow as they
fall. The growth is well fit with a gravitational stretching
function which indicates that perturbations of the order of a grain
size at the nozzle serve as nucleation points for the drops. At
higher pressures the jet atomizes below some depth. The depth at
which disintegration occurs increases with decreasing pressure. At
the lowest available pressure, $p\approx 1/5000^{\textnormal{th}}$
atm, the jet does not disintegrate within the experimentally
accessible range of depth.

The following section gives an overview of inhomogeneities that can
arise in granular flows. Section III describes the experimental
method. The next section contains the experimental results and is
followed by a discussion and conclusion in sections V and VI.

\section{II. Background}
Intermittency and clogging are intrinsic features of granular flows.
This is due to friction between grains and geometric constraints
that prevent grains from flowing past each other when the flow is
confined by boundaries (jamming). For finer particles air can also
induce intermittency. This is observed in various systems, such as
flow down a vertical pipe or in hourglasses
\cite{schick,bertho,raafat,wu,veje1,veje2}. In my experiment the
system undergoes a change from a dense granular funnel flow to a
freely falling flow. Both regimes are susceptible to intermittency
and clustering. In the following, different mechanisms that create
inhomogeneities in granular flows and gases are reviewed.

\subsection{A. Density waves}
Density fluctuations can propagate through a flow, such as in
vertical pipes \cite{schick,bertho} or funnel flow. Granular funnel
flow and the related hopper flow have been extensively studied in
the literature. The flow out of a nozzle can become inhomogeneous in
a 2D as well as a 3D funnel system \cite{brown,baxter}. As the
grains move downward under the influence of gravity, arches form
near the nozzle that can temporarily hold up the flow before they
collapse. The presence of these density waves is enhanced when the
grains are rough and is suppressed when the grains are smooth
spheres \cite{baxter}.

\subsection{B. Interstitial fluid effects}
Apart from grain-grain and grain-boundary interactions, interstitial
fluid such as air can profoundly affect the flow. This is especially
important for systems such as mine where the particle size is well
below $1$ mm \cite{duran}. In this regime hydrodynamic forces can
influence the dynamics strongly via viscous drag and pressure
gradients building up inside the bed. The latter occurs because the
movement of air is impeded inside a porous medium. In typical
experimental settings ($v\lesssim 1$ m/s, $p=1$ atm) this effect
does not matter much for grain sizes $\gtrsim 1$ mm, but becomes
important for smaller media. As a result, the granular flow out of
an aperture can become oscillatory \cite{wu,veje1,veje2}. This
so-called ``ticking'' is due to pressure gradients inside the bed
that create a back-flow of air into the nozzle.

In cases where the interstitial fluid is a liquid, interesting flow
instabilities have been observed \cite{nicolas, schaflinger}.
Nicolas \cite{nicolas} studied a system similar to mine, namely a
granular jet emanating from a nozzle, but in the presence of a
liquid. Depending on the grain and liquid parameters, the jet can
either remain homogeneous or become unstable and form blobs similar
to the clusters in my experiment. The origin of this instability
still remains unclear. Nicolas showed that treating the suspension
jet as a fluid with some effective viscosity cannot explain this
instability.

\subsection{C. Inelastic Clustering}
A granular gas is dissipative due to the inelastic nature of the
collisions between the grains. Therefore, without external energy
supply the gas cools down with time and eventually freezes
\cite{mcnamara,goldhirsch,poeschel}. However, it does not cool
homogeneously, but forms clusters. When a fluctuation increases the
density locally, the collision rate goes up in this region. Due to
the increased dissipation, the granular temperature decreases which
in turn lowers the pressure. The resulting pressure gradient will
enhance migration of particles into that region, thereby increasing
the density even further. Eventually, these regions grow into
clusters.
\subsection{D. Cohesion}
Humidity in the air and surface charges can induce cohesive forces
between particles \cite{duran}. Condensates from ambient humidity
create liquid bridges causing particles to stick together. As with
interstitial fluid effects, the influence of cohesion depends
strongly on the size and the density of the particles. Smaller and
lighter media are more susceptible to cohesive forces. Cohesion does
not cause intermittency per se, but can dramatically change the
rheological properties of granular media. In extreme cases it leads
to clumping and caking. These effects can be controlled but not
completely eliminated. In this experiment, I keep the relative
humidity at $50\%$, which provides enough ions to eliminate surface
charges, but does not result in clumping. However, small residual
cohesive forces could potentially induce clustering in the freely
falling jet.

\section{III. Experimental Method}
In this paper I exclusively study $d=100$ $\mu\textnormal{m}$
spherical glass beads (Mo-Sci Corp., $\rho=2500$ g/l) (Fig.
\ref{100micron}). In order to investigate the clustering in the
emerging jet quantitatively, an optical system was set up to measure
inhomogeneities in the jet - similar to ones used in previous
studies for measuring intermittent granular flows in hourglasses and
vertical pipes \cite{schick,raafat,veje1}. The basic idea is to
measure the light intensity of a laser shining through the falling
grains. The fluctuations in the intensity allow me to record
inhomogeneities in the flow. This method does not measure density
variations, but rather the emergence of undulations on the jet
surface and gaps.

\begin{figure}
\begin{center}
\includegraphics[width=3.4in]{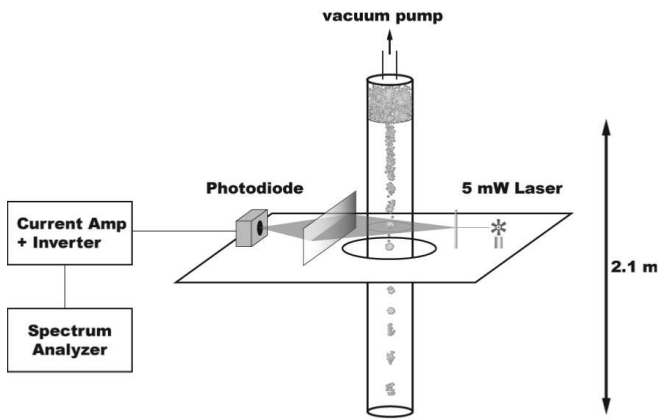}
\caption{\label{setup} Setup. The granular jet emerges from the
reservoir at the top of the $14$ cm wide acrylic tube. The tube is
connected to a vacuum pump and the pressure is measured with a
pressure gauge mounted on the lid. The optical platform can be
adjusted to any height between the nozzle and $2.1$ m below. The
light from a $5$ mW ($\lambda=650$ nm) laser diode is focused by a
lens, then passed through a glass rod to spread it out and finally
refocussed with a cylindrical lens onto a photodiode. An identical
laser sheet (not shown) is mounted $5.1$ mm below the first sheet.
The signals of the two photodiodes are passed to two current amps
that also invert the signals. The two signals are cross-correlated
with a spectrum analyzer and one of them is sent to an A/D converter
card.}
\end{center}
\end{figure}

The schematics of the setup are shown in Fig. \ref{setup}. The
funnel is mounted inside and near the top of a $14$ cm wide acrylic
tube whose diameter is large enough so that the particles are always
far from the wall. A $9$ cm wide cylindrical reservoir feeds the
grains into a porous tube. At the bottom of this tube a disc with a
$4$ mm circular aperture is attached from which the particles emerge
and then freely fall down the tube. There is a remote controlled
shutter beneath the nozzle to initiate and stop the flow. The
pressure inside the acrylic tube can be pumped down to $0.02$ kPa.
For air, this corresponds to a mean free path of $0.3$ mm ($\approx
3$ d) \cite{chemistry}. The pressure is monitored with a pressure
gauge mounted on the lid (Granville-Phillips, Convectron gauge 375).
An oil filter (K.J. Lesker micromaze foreline trap) and a shut-off
valve was installed between the system and the vacuum pump to avoid
contamination of the system with oil vapor that might emanate from
the vacuum pump.

\begin{figure}
\begin{center}
\includegraphics[width=2in]{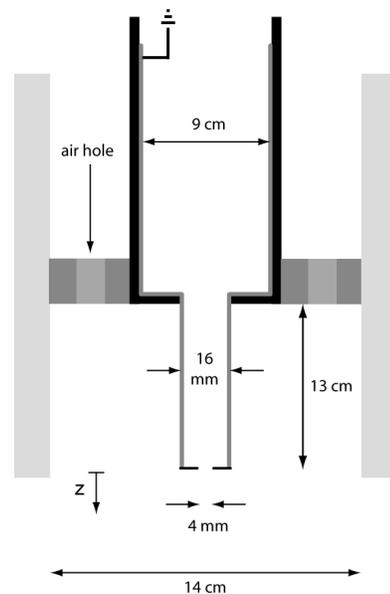}
\caption{\label{reservoir} Schematics of the reservoir. The
reservoir consists of a $9$ cm wide acrylic tube which feeds a $16$
mm wide porous tube. At the bottom of the porous tube I attached a
metal disc with a $4$ mm circular aperture. The inside is covered
with metal mesh and is grounded. The reservoir sits on a ring that
has several holes in it to allow for pressure equalization between
the reservoir and the lower part of the tube. The depth $z$ is
measured from the nozzle.}
\end{center}
\end{figure}

\begin{figure}
\begin{center}
\includegraphics[width=3.4in]{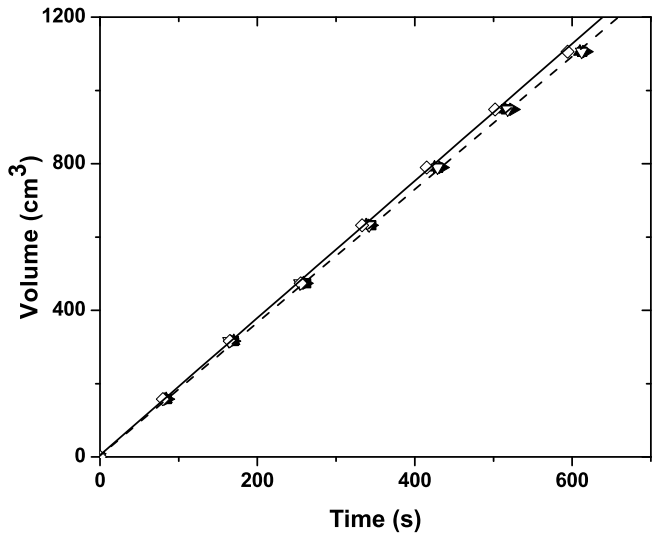}
\caption{\label{flowrate} Volume flow vs. time. The solid and dashed
lines are linear fits with respective slopes of $1.87$ and $1.82$
$\textnormal{cm}^3/\textnormal{s}$. $(\blacksquare)$, $p=101$ kPa;
$(\blacktriangleright)$, $p=62$ kPa; $(\blacktriangle)$, $p=26$ kPa;
$(\triangledown)$, $p=3$ kPa; $(\lozenge)$, $p=0.04$ kPa.}
\end{center}
\end{figure}

A prerequisite for this experiment is to ensure a steady
non-changing flow out of the nozzle. It is known that air pressure
gradients across the bed can cause oscillatory flow ("ticking") out
of the nozzle \cite{wu}. Even equalizing the pressure in the
reservoir with the pressure at the nozzle is not sufficient for
steady flow conditions. To eliminate ticking, I made the tube coming
from the reservoir porous. It is made out of metal mesh ($0.0014''$
wire mesh $325$x$325$) that is permeable to air, but not to the
particles, thereby allowing pressure equalization through the
boundaries (Fig. \ref{reservoir}). This ensures a stable flow rate
at all pressures (Fig. \ref{flowrate}).

The whole optical system is mounted on a platform that is moveable
in the vertical direction and can be lowered to $2.1$ m below the
nozzle.  Since the width of the freely falling particle flow exceeds
the diameter of the laser beam (5 mW laser diodes from z-bolt.com),
the light is spread out to a sheet to capture the entire horizontal
spread of the flow. This is achieved by shining the light through a
$8$ mm diameter glass rod. The beam is also focused in the
horizontal plane by a lens to ensure that the beam waist is smaller
than the emerging structures. The vertical width is $0.3$ mm
($\approx 3d$). After the laser sheet passes through the particle
flow, it is refocused by a cylindrical lens onto a photodiode
(Silonex SLD-68HL1D) that measures the intensity. A current amp is
used to convert the diode current, which is linearly proportional to
the incident light intensity, into a voltage signal. The signal is
inverted since the quantity of interest is the amount of light that
is screened by the jet. We refer to this signal as the blockage $B$.
The frequency response is flat up to at least $100$ kHz. The signal
is recorded with an A/D converter card at a sampling rate of $48$
kHz. The baseline of the signal changes with vertical position due
to imperfections of the acrylic tube. The signal is scaled with the
baseline at each position before each experimental run.

In order to convert time scales into length scales, the local
average velocity of the flow is measured by cross-correlating the
intensity signal of two closely spaced laser sheets (vertical
distance$=5.1$ mm). In order to avoid cross-talk between the two
photodiodes through scattered light a shutter is mounted near the
cylindrical lens. A $2$-channel spectrum analyzer (SRS SR780)
obtains the cross correlation between the two signals in real time.

When calculating autocorrelations, the mean of the signal is
subtracted and the autocorrelation is normalized to $1$ at zero time
delay $\Delta t$. In that way, the autocorrelation approaches zero
at large $\Delta t$. Near the nozzle, the signal to noise is low and
a noise floor appears in the autocorrelation. The electric noise
floor is measured directly by obtaining the autocorrelation of the
laser signal without the granular jet. It is then subtracted from
the autocorrelation.

Cohesion between particles is caused by liquid bridges and/or
electrostatic interaction due to charge buildup \cite{duran}. The
latter is especially significant in dry atmospheres. To ensure
stable conditions the laboratory is controlled at $50\%$ relative
humidity. Moreover, the walls of the reservoir are grounded to avoid
build up of charge through friction. After each run, the beads are
exposed to ionized air to neutralize any charges that might have
built up during the run.

\section{IV. Results}
In order to compare the clustering at different pressures we need to
know the flow rate out of the nozzle and the velocity of the jet as
a function of pressure. As mentioned above, a porous tube is used to
ensure steady flow at higher pressures. Furthermore, care has been
taken to prepare the glass spheres in the reservoir in the same way
after each run so that the packing fraction remains constant. Figure
\ref{flowrate} shows the volume flow rate at different pressures.
This was measured by recording how long it takes for the full
reservoir to drain $158$ $\textnormal{cm}^{3}/\textnormal{s}$ at a
time. I do not let the reservoir empty completely, since the flow
rate changes when the fill height is low. The flow rate is nearly
constant for all pressures: At atmospheric pressures the flow rate
is $1.82$ $\textnormal{cm}^{3}/\textnormal{s}$, which is slightly
lower than at $p=0.04$ kPa, where it is $1.87$
$\textnormal{cm}^{3}/\textnormal{s}$. This few percent difference is
negligible for the subsequent analysis. The constant flow rate
allows a direct comparison of the clustering at different pressures.

\begin{figure}
\begin{center}
\includegraphics[width=3.4in]{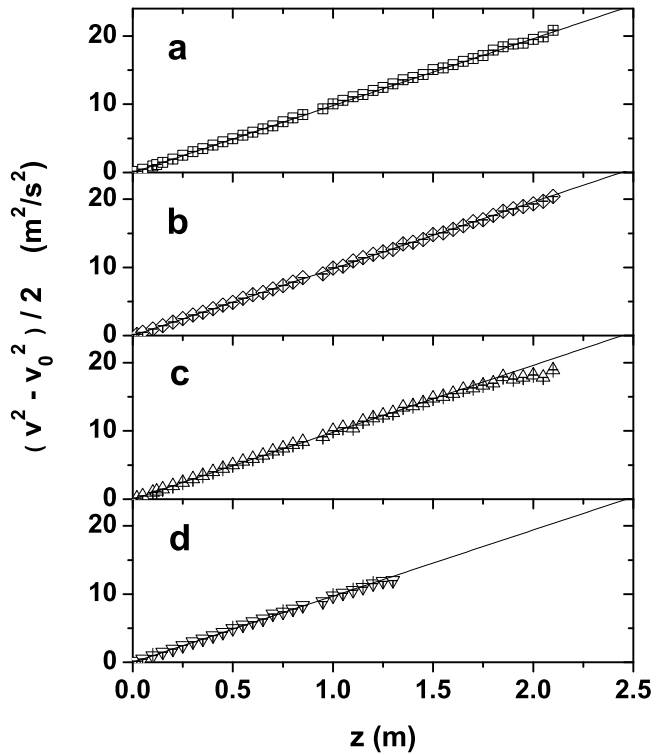}
\caption{\label{velplot} Velocity profile of jet at different
pressures. In all panels, $v_{0}=0.36$ m/s. The solid lines are
linear fits to the data. The slope represents the acceleration $a$.
(a) $p=0.027$ kPa, $a=9.8$ m/s; (b) $p=0.67 kPa$, $a=9.8$ m/s;(c)
$p=49$ kPa, $a=9.8$ m/s;(d) $p=101$ kPa, $a=9.7$ m/s.}
\end{center}
\end{figure}

In order to convert the signals recorded in the time domain into the
length domain, we need to know the velocity at each height. Figure
\ref{velplot} shows the velocity of the jet versus depth at four
pressures. In the absence of any hydrodynamic drag force the
velocity $v$ should just follow $v=\sqrt{2gz+v_{0}^2}$, where $z$ is
the distance as measured from the aperture, $v_{0}$ the velocity at
the nozzle and $g$ is the acceleration of gravity. When
$(v^{2}-v_{0}^{2})/2$ is plotted against the depth $z$, the
resulting curve is linear for simple free fall with a slope equal to
the acceleration $a=g$. $v_{0}$ is constant within a few percent at
all pressures. This is consistent with the constant flow rate found
earlier. Down to the lowest available pressure we find a linear
relationship between $(v^{2}-v_{0}^{2})/2$ and $z$ with a slope
close to or equal to $g$. At higher pressures deviations are
observed when the jet starts to disintegrate into a cloud of
particles (Fig. \ref{100micron} (e)). When $p=101$ kPa ($=1$ atm),
this happens around $1.2$ m below the nozzle (Fig. \ref{velplot}
(d)). At this point the mean velocity stops growing with depth and
the previously sharp cross-correlation peak becomes broad. When the
pressure is lowered, the disintegration starts further downstream
(Fig. \ref{velplot} (c)). At the lowest pressure (Fig. \ref{velplot}
(a)), no disintegration is observed in the experimentally available
range of depth.

The low drag on the jet at $p=101$ kPa is surprising given that the
viscous drag on a single grain is substantial and would lead to a
terminal velocity of $0.76$ m/s according to the Stokes formula for
viscous drag on a sphere: $F_{\textnormal{Stokes}}=3\pi \mu d v$. A
possible explanation is that the air is essentially trapped inside
the jet so that it appears solid. Treating the jet as a porous
medium, one can estimate the time it takes for air to penetrate the
jet. Inside a porous medium, the pressure obeys a diffusion equation
\cite{carman}. The diffusion constant is $D=(P_{0}k)/(\mu(1-\phi))$,
where $P_{0}$ is the ambient gas pressure, $\mu$ the dynamic
viscosity of air, and $k$ the permeability of the granular medium at
a packing fraction $\phi$. The permeability is an empirical constant
that is well approximated by the Carman-Kozeny relation
\cite{carman}: $k=d^2(1-\phi)^3/(180\phi^2)$. The value for $D$ is
$0.3$ $\textnormal{m}^2/\textnormal{s}$ after substituting numerical
values for the constants: $\mu=1.8 \cdot 10^{-5}$ Pa s, $P_{0}=101$
kPa and $\phi=0.5$. The latter is a typical packing fraction for a
random loose pack of spheres. This value decreases as the jet falls
and gets stretched. The typical time $\tau$ for air to diffuse into
the jet is therefore $\tau \approx r_{jet}^2/(2D) = 6.7$ $\mu
\textnormal{s}$. This is small compared to other typical timescales
of the system, such as the time it takes for a grain to fall its own
diameter: $d/v \approx 10^{-4}$ s. This means that the jet is
permeable to air. Therefore, a more sophisticated hydrodynamic
description is needed to explain the low drag on the jet at
atmospheric pressure.

\begin{figure}
\begin{center}
\includegraphics[width=3.4in]{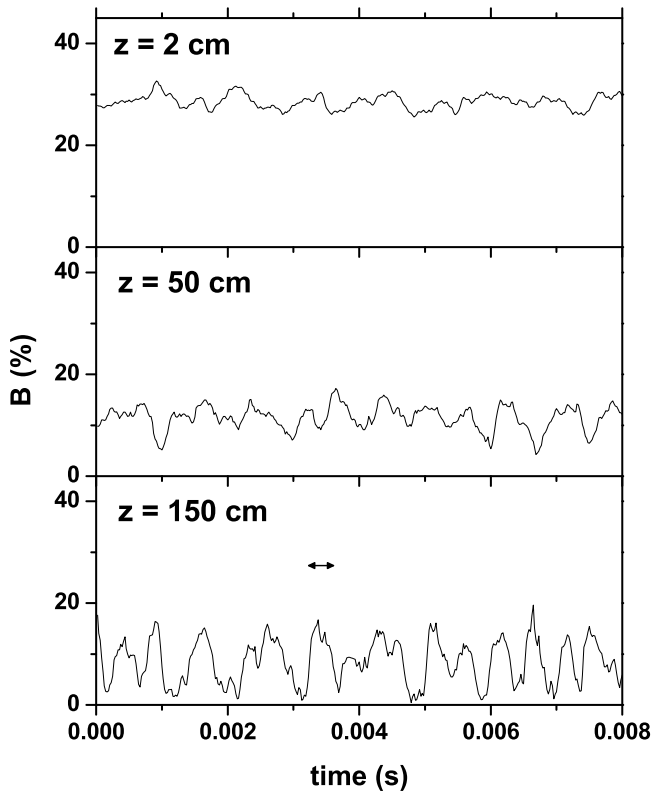}
\caption{\label{tracevac} Blockage at different depths, $z$, at
$p=0.027$ kPa. The length of the arrow equals the dip position of
the corresponding autocorrelation.}
\end{center}
\end{figure}

\begin{figure}
\begin{center}
\includegraphics[width=3.4in]{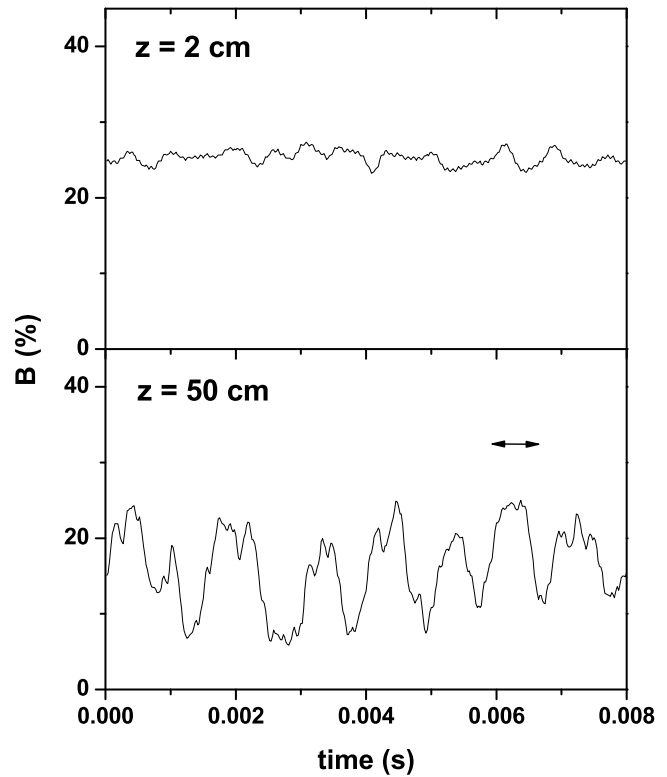}
\caption{\label{traceair} Blockage at different depths, $z$, at
$p=101$ kPa. The length of the arrow equals the dip position of the
corresponding autocorrelation.}
\end{center}
\end{figure}

Figures \ref{tracevac} and \ref{traceair} display typical time
traces of the signal at $p=0.027$ kPa and atmospheric pressure,
respectively. Just below the nozzle, $z=2$ cm, drops have not yet
formed and the signal fluctuates only slightly. At $z=50$ cm the
fluctuations have visibly increased and finally, at $z=150$ cm, the
signal contains clear peaks which show the presence of well defined
drops.
At atmospheric pressure the behavior is similar. Comparing the time
traces at $z=50$ cm, it is apparent that the drops are bigger at
atmospheric pressure. In atmosphere only depths up to $\approx 120$
cm can be probed, since the jet disintegrates beyond that depth.

\begin{figure}
\begin{center}
\includegraphics[width=3.4in]{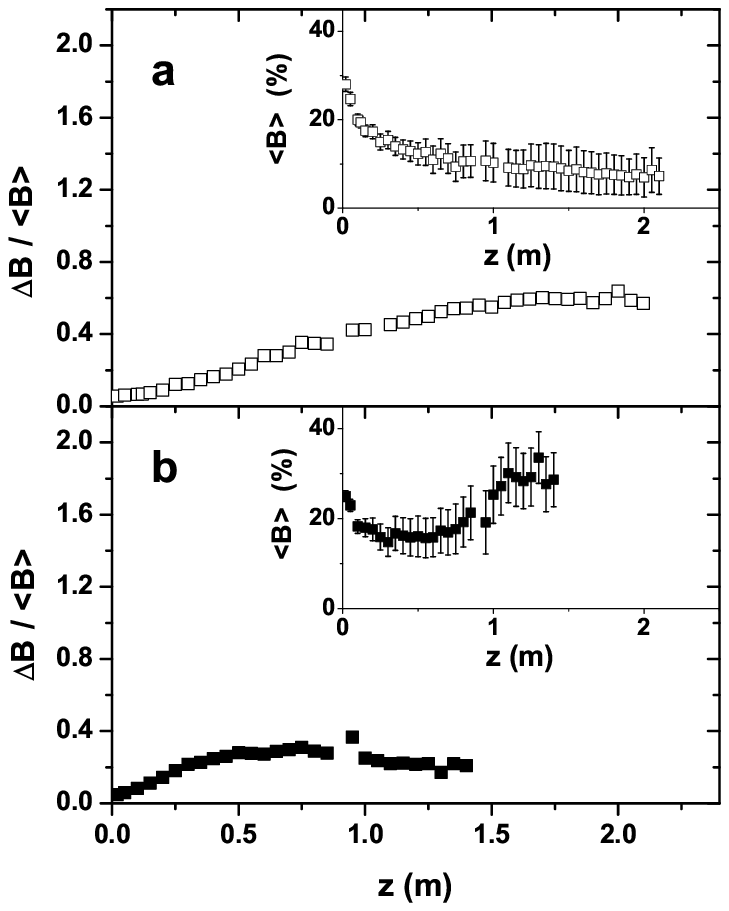}
\caption{\label{fluct} Blockage fluctuations as a function of depth.
(a) $p=0.027$ kPa, (b) $p=101$ kPa. Insets: Mean and standard
deviation of the blockage as a function of depth at the respective
pressure. The error bar delineates the standard deviation $\Delta
B$.}
\end{center}
\end{figure}

In order to illustrate the evolution of the drops, the fluctuations
of the blockage signal as a function of depth have been plotted
(Fig. \ref{fluct}). The fluctuations are just the standard
deviation, $\Delta B$, divided by the mean of the signal $<B>$. In
vacuum (Fig. \ref{fluct} (a)) the fluctuations increase
monotonically with depth and then saturate. Similarly, at
atmospheric pressure the fluctuations increase, but then slowly
decrease (Fig. \ref{fluct} (b)).

The insets in Fig. \ref{fluct} show how $<B>$ and $\Delta B$ vary
with depth. At low pressure, the mean decreases with depth while the
standard deviation increases. This reflects the increase of
undulations and gaps with increasing depth. At atmospheric pressure
(Fig. \ref{fluct} (b)), the mean varies non-monotonically. Beyond
$1$ m it even exceeds the value at small depth when the jet is still
compact. The standard deviation, however, stops increasing far away
from the nozzle.

The reason for this marked difference in behavior is that in the
presence of air, the jet does not stay collimated far away from the
nozzle as it does at low pressures. This leads to an increased
blockage as the jet starts to spread. Furthermore, there is more
spray, presumably caused by advection of particles by the
surrounding air. The optical signal is sensitive to this spray since
it scatters the light. This explains why the mean blockage starts to
rise again below $60$ cm or so. It should also be noted that even if
the jet is still homogeneous, the blockage does not reach $100\%$ at
both pressures, since the laser sheet is wider than the jet
diameter.

\begin{figure}
\begin{center}
\includegraphics[width=3.4in]{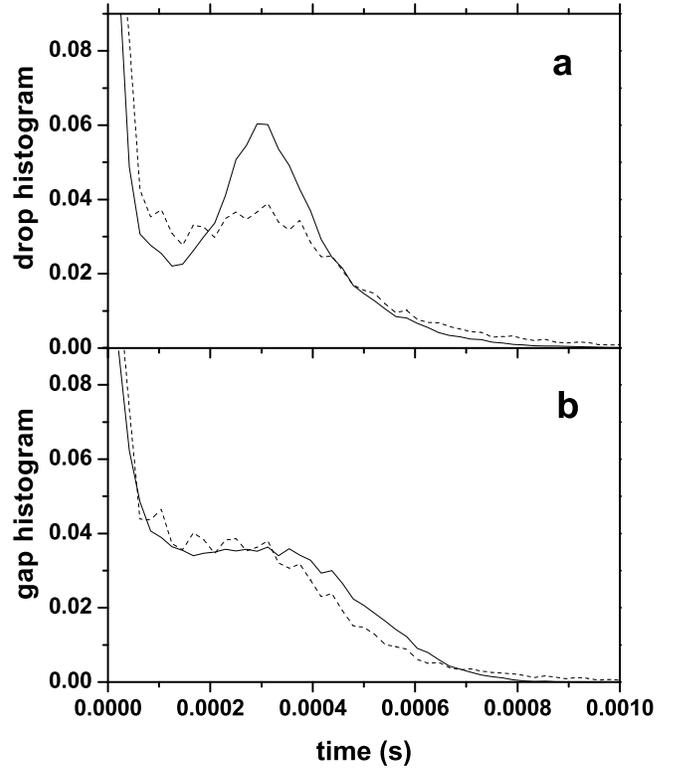}
\caption{\label{histo} Histogram of drops and gaps at $p=0.027$ kPa.
In both panels the solid line corresponds to $z=150$ cm and the
dashed line to $z=50$ cm. The threshold to discriminate between drop
and gaps is the average of the signal. (a) drop histogram; (b) gap
histogram}
\end{center}
\end{figure}

One way to analyze the signal is to choose a threshold and convert
the blockage signal into a binary sequence as done by Raafat and
coworkers \cite{raafat}. Anything above the threshold is considered
a drop, anything below a gap. The resulting histogram for my
experiment is shown in Fig.\ref{histo}. This was obtained at
$p=0.027$ kPa for two heights and the threshold was taken to be the
signal average. The drop histogram displays a clear peak at $z=150$
cm, while it is less pronounced at $z=50$ cm. The gap histogram
remains flat and falls off at large gap sizes in both cases. The
high occurrence of very small drop and gap sizes is due to
electronic noise and particle spray. The histograms bear some
resemblance to the ones in the vertical pipe flow studied by Raafat
et al. \cite{raafat}. Fig. \ref{histo} shows the emergence of a
typical drop size, while the gap size distribution is broad.

\begin{figure}
\begin{center}
\includegraphics[width=3.4in]{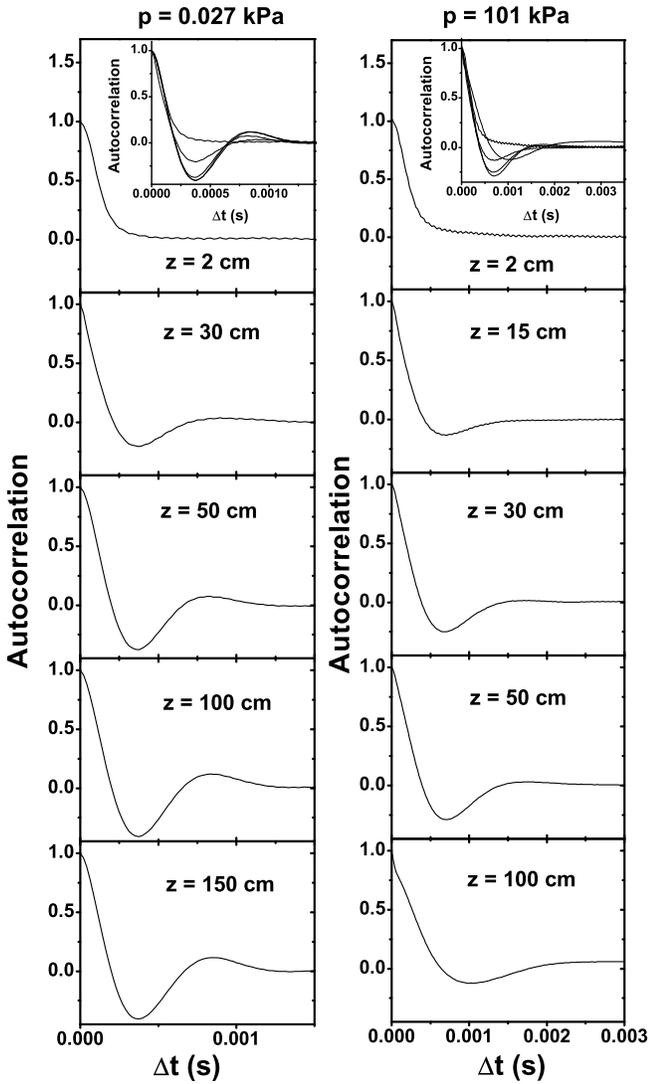}
\caption{\label{acorr} Autocorrelation at different heights. Left
panels show data at $p=0.027$ kPa, the right panels at $p=101$ kPa.
The depth at which the signal was recorded is indicated in each
panel. The insets in the upper panels show all the autocorrelations
for the respective pressure.}
\end{center}
\end{figure}

This analysis is prone to noise. The peak in the drop histogram
appears only below $50$ cm, even though inhomogeneities appear
earlier (by visual inspection with a stroboscope). A more sensitive
analysis is the autocorrelation as shown in Fig.\ref{acorr}.

Autocorrelations are displayed for two different pressures. At both
pressures, a dip and a peak in the autocorrelation develops, though
the peak is far less pronounced at $p=101$ kPa. The dip position
does not change with depth at $p=0.027$ kPa. At atmospheric
pressure, the dip remains constant as a function of depth initially,
but then moves to higher $\Delta t$ below $z=80$ cm. Also, the dip
position at $p=101$ kPa is significantly larger than at low
pressure. At low depths ($z<80$ cm) it is $7.0 \cdot 10^{4}$ s
compared to $3.8 \cdot 10^{-4}$ s at $p=0.027$ kPa.

A priori it is not clear whether the dip represents the time scale
of drops or gaps. Looking at the time traces in Figs. \ref{tracevac}
and \ref{traceair}, drops and gaps have similar sizes. The histogram
(Fig. \ref{histo}) confirms this observation. The drop and gap size
distribution are both broad and fall off around $0.008$ s.
Therefore, the dip reflects the typical size of the fluctuations in
the time domain. The length of the arrows in Figs. \ref{tracevac}
and \ref{traceair} represent the dips in the respective
autocorrelations of the time traces. They show good agreement with
the typical fluctuation size by visual inspection.

Summarizing the above, we identify the dip position with the typical
fluctuation size (in the time domain) and find that with increasing
depth it stays constant for low pressures and becomes larger for
atmospheric pressure. It follows that the clusters and gaps
continually grow as they are stretched by the increasing velocity as
a function of depth.

In order to understand the growth of the drops it is instructive to
consider the case where the jet is just stretched due to gravity
ignoring all other interactions. The equations of motion for two
grains, one right at the nozzle starting with velocity $v_{0}$, the
other a distance $\lambda_{0}$ below, are respectively:
\begin{eqnarray}
z &=& v_{0}t+\frac{1}{2}gt^{2}\\
z' &=& \lambda_{0}+v_{1}t+\frac{1}{2}gt^2
\end{eqnarray}
where $v_{1}=\sqrt{2g\lambda_{0}+v_{0}^2}$ is the velocity the grain
attains by falling a distance $\lambda_{0}$. Due to gravitational
acceleration the initial grain separation $\lambda_{0}$ will grow in
time as $\lambda(t)=z'-z$.

Parameterizing $t$ with $z$ we obtain
\begin{equation}\label{stretch}
\lambda(z)=\lambda_{0}+\frac{v_{0}^2}{g}
\left(\sqrt{\frac{2g\lambda_{0}}{v_{0}^2}+1}-1\right)\left(\sqrt{\frac{2gz}{v_{0}^2}+1}-1\right).
\end{equation}
When $\lambda_{0}\ll \frac{v_{0}^2}{2g}$ this reduces to
\begin{equation}
\lambda(z)=\lambda_{0}\sqrt{\frac{2gz}{v_{0}^2}+1}.
\end{equation}
Far away from the nozzle, when $z\gg \frac{v_{0}^2}{2g}$, this can
be further simplified to
\begin{equation}\label{stretch1}
\lambda(z)=\lambda_{0}\frac{\sqrt{2gz}}{v_{0}}.
\end{equation}
Since $\sqrt{2gz}$ is just the velocity $v(z)$ at depth $z$ for
$z\gg \frac{v_{0}^2}{2g}$, equation \ref{stretch1} can be written as
\begin{equation}\label{stretch2}
\frac{\lambda_{0}}{v_{0}}=\frac{\lambda(z)}{v(z)} \equiv T,
\end{equation}
where $T$ is the time it takes to fall a distance $\lambda(z)$ at
depth $z$. This time is constant for $z\gg \frac{v_{0}^2}{2g}$. We
can now compare $T$ with the dip position of the autocorrelation
which is a measure of how long it takes for a drop to fall. At
$p=0.027$, the dip position is constant at all depths to within a
few percent: $3.8\cdot 10^{-4}$ s; the clusters just get stretched
by falling in gravity. The velocity at the nozzle is $v_{0}=0.36$
m/s. Therefore, $\lambda_{0}=1.4 \cdot 10^{-4}\textnormal{m}=1.4 d$
(this also justifies the approximation $\lambda_{0}\ll
\frac{v_{0}^2}{2g}$ to simplify equation \ref{stretch}). At
atmospheric pressure the dip position is larger and grows below
$z=80$ cm. Using the value of the dip position for smaller depths,
we find $\lambda_{0}=2.5\cdot 10^{-4}\textnormal{m}=2.5 d$.

\begin{figure}
\begin{center}
\includegraphics[width=3.4in]{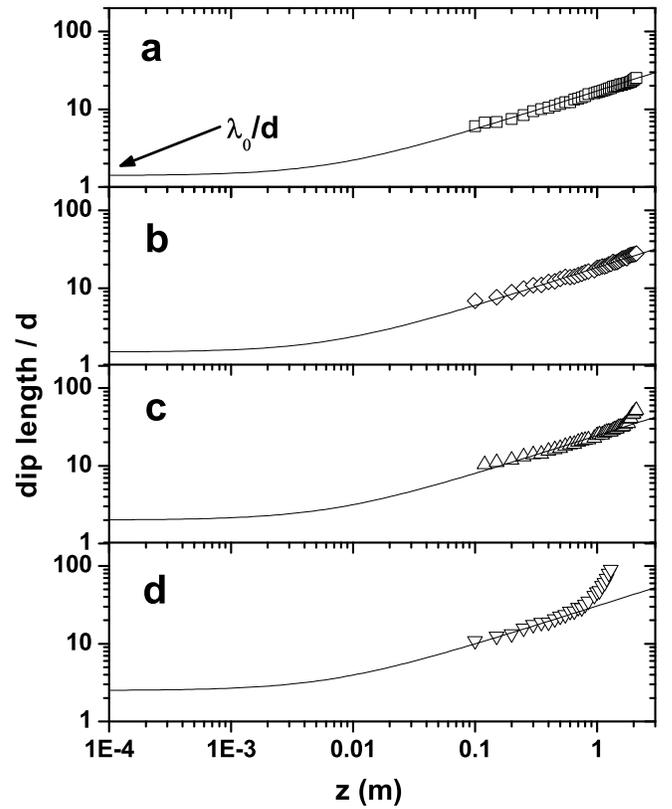}
\caption{\label{dippress} Dip length vs. depth at different
pressures. Each dip position has been converted into a length scale
by multiplying it with the local velocity. The solid lines are fits
from equation \ref{stretch}. The error bars are the size of the
symbols. (a) $p=0.027$ kPa; (b) $p=0.67 kPa$; (c) $p=49$ kPa; (d)
$p=101$ kPa.}
\end{center}
\end{figure}

In order to see how the drop growth compares with pure gravitational
stretching, I plot the dip length as a function of depth measured at
four different pressures (Fig.\ref{dippress}). The dip positions
have been converted into length scales by multiplying each of them
with the local velocity that is obtained from Fig. \ref{velplot}. I
plot the gravitational stretch equation (eqn. \ref{stretch}) for
each data set using the value for $\lambda_{0}$ we found from our
previous considerations, so there are no free fitting parameters.
The dip lengths have been rescaled with the particle diameter. At
$p=0.027$ kPa, the data is well fit by this equation. Unfortunately,
the optical setup is not sensitive enough to pick up inhomogeneities
above $10$ cm depth. The fit for higher pressures agrees with the
data until the onset of atomization. At that point, the dip position
grows larger than it would just with gravitational stretching.

\begin{figure}
\begin{center}
\includegraphics[width=3.4in]{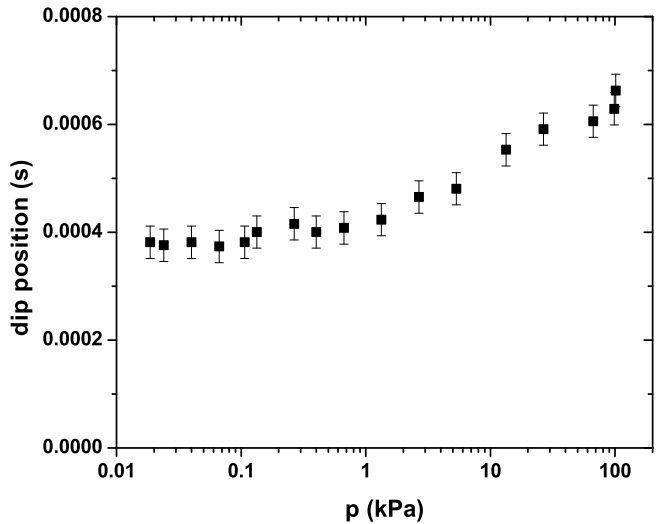}
\caption{\label{pressplot} Dip position as a function of pressure at
$z=20$ cm.}
\end{center}
\end{figure}

In  Fig. \ref{pressplot} I track the dip position in the
autocorrelation as a function of pressure at constant depth $z=20$
cm below the nozzle. The dip position decreases with decreasing
pressure until about $p\approx 0.1$ kPa at which it stays constant
down to the lowest available pressure $0.02$ kPa. As we have seen
before, the air has an appreciable effect on the drop size. The dip
changes by almost a factor of two. Moreover, it shows that air has
no effect below $0.1$ kPa. Below that pressure the dip position
remains constant over one order of magnitude in pressure.

\section{V.Discussion}
I have shown that the clustering of a freely falling granular jet is
influenced by the presence of air, but occurs even at low pressures.
At the lowest available pressure, $p=0.02$ kPa, the mean free path
exceeds the particle size by a factor of $3$, so air effects should
be negligible. Moreover, at constant depth, the position of the
first dip in the autocorrelation function, which is a measure of the
typical size of the structure, stays constant below $0.1$ kPa down
to $0.02$ kPa (Fig.\ref{pressplot}). Therefore, air-grain
interactions can be ruled out as the initial cause of the
clustering.

It is worth noting that other granular phenomena that depend on the
ambient pressure lose their pressure dependence below $\approx 0.1$
kPa. Granular size separation in vibrated beds depends strongly on
pressure when the grain size is small ($d\lesssim
500\mu$m)\cite{mobius, king}. It has been found that below $0.1$ kPa
air ceases to play a role. Another granular phenomenon that depends
on air is heaping. The surface of a bed of small particles
($d\lesssim600\mu\textnormal{m}$) starts to tilt when vertically
vibrated \cite{pak}. It was found that heaping dramatically
decreases below $1$ kPa, since the mean free path of air becomes
comparable to the grain size. This suggests that it is for this same
reason that the dip position stops changing at low pressures in
Fig.\ref{pressplot}.

This experiment does not measure density, so I cannot distinguish
between cluster growth through agglomeration and the stretching of a
cluster due to gravity. Nevertheless, the latter is always present,
so it is sensible to compare the growth in cluster size with
gravitational stretching.

My results indicate that the instability arises from fluctuations on
the granular level at the nozzle. At the lowest available vacuum,
where air effects are negligible, gravitation is the only external
force acting on the particles. Indeed, the fit for gravitational
stretching is good at these low pressures. Extrapolating the drop
growth back to the nozzle, I find the initial size to be of the
order of a grain diameter.

At higher pressures deviations from the gravitational stretching are
observed. The extrapolation yields an initial size that is almost a
factor of two larger than at low pressures. Moreover, the jet starts
to disintegrate into a cloud below some depth. This is presumably
due to hydrodynamic interactions, since we do not observe this at
low pressures, at least in the observable range of depths.

I did not observe clustering for grain sizes larger than $200\mu
\textnormal{m}$. Visual inspection by strobing granular jets of
larger particles did not reveal any inhomogeneities visible by eye.
Smaller grains, on the other hand, give rise to strong clustering.

It is still unclear how these grain-sized fluctuations grow into
drops of several particles. The fact that clustering is not observed
for glass spheres $\gtrsim 200 \mu\textnormal{m}$ suggests that
cohesive forces might be responsible. However, if cohesive forces
were dominant, so that the particles just stick together, the drops
should not grow and remain constant in size once they are formed. On
the other hand, the velocity correlations induced by inelastic
collisions of equal particles do not depend on the mass of the
particles and therefore should not depend on their size. They just
depend on the coefficient of restitution. However, there are
indications \cite{luding, labous} that the coefficient of
restitution is size dependent. Experimentally \cite{labous} and
theoretically \cite{luding} it was shown that smaller spheres give
rise to a smaller coefficient of restitution. Therefore, a granular
gas of small spheres should form clusters more rapidly than larger
ones. Since I do not have any information on the coefficient of
restitution of the glass spheres used in my experiment, it is
unclear whether this size dependence could explain the absence of
clusters in granular jets of large particles.

Regardless of the clustering mechanism, an initial perturbation is
required that acts as a seed for the cluster. Since gravity cannot
initiate fluctuations and air-grain interactions have been ruled
out, this perturbation is likely to come from the nozzle. This is
consistent with the finding that the cluster size extrapolates to a
few grain diameters at the nozzle, which is a typical fluctuation in
granular flows.

\section{VI. Conclusion}
In contrast to ordinary fluids, granular flows lack surface tension
and are discrete in nature. Despite these fundamental differences, a
freely falling jet becomes unstable and forms drops in both cases.
However, the physical origin of the instability in the granular case
is quite different from its fluid counterpart.

The length scale, after tracing it back to the nozzle using the
gravitational stretch equation, turns out to be of the order of a
grain size at all pressures. This suggests that granular
fluctuations at the nozzle set the cluster size. Moreover, I have
shown that the surrounding liquid, in my case air, is not required
for drop formation. It does however, change the length scales by
almost a factor of $2$ and ultimately leads to the atomization of
the jet. The latter is not observed at low pressures.

These results might have implications on the clustering phenomenon
found in suspensions \cite{nicolas, schaflinger}. They have
typically been ascribed to an effective surface tension between the
fluid and the sediment. My results show that fluctuations on the
granular level can propagate downstream and thereby impose a length
scale. This suggests that the granular nature of the sediment cannot
be ignored and a continuous medium description is inadequate to
explain this phenomenon.

This study also leaves some open questions: Clustering is not
observed for glass spheres larger than $200 \mu$m. This observation
should give insight into the clustering mechanism which still
remains unclear. How do fluctuations of single grains grow into
clusters of many particles? Further studies are needed to answer
this question.

The clustering of a freely falling granular jet is a novel granular
instability that is initiated by fluctuations at the nozzle that are
of the order of a grain diameter. These fluctuations grow into
clusters downstream. The detailed structure of the drops may reveal
information about these fluctuations that are experimentally
difficult to access otherwise.

\section{Acknowledgements}
I thank H. Jaeger and S. Nagel for their support and guidance. I am
grateful to X. Cheng, E. Corwin, M. Holum and J. Royer for useful
discussions and suggestions. H. Krebs's expertise was invaluable to
build the apparatus. This work is supported by MRSEC, No.
DMR-0213745.

\end{document}